\documentclass[12pt]{article}
\usepackage[utf8]{inputenc}
\usepackage{amsmath, amssymb, url, float, graphicx, float}
\usepackage{fullpage}
\usepackage[small,bf]{caption}
\usepackage{cite}
\usepackage{xcolor}
\usepackage{nicefrac}

\usepackage{amsthm}
\theoremstyle{definition}

\newtheorem{theorem}{Theorem}
\newtheorem{claim}{Claim}

\counterwithin*{lemma}{subsection}

\let\phi\varphi

\newcommand{\ones}{\mathbf 1}
\newcommand{\reals}{{\mbox{\bf R}}}

\newcommand{\naturals}{{\mbox{\bf N}}}

\newcommand{\Ttr}{\mathop{\bf Tr}}

\newcommand{\Expect}{\mathop{\bf E{}}}
\newcommand{\Var}{\mathop{\bf Var{}}}

\newcommand{\eg}{{\it e.g.}}
\newcommand{\ie}{{\it i.e.}}

\newcommand{\BEAS}{\begin{eqnarray*}}
\newcommand{\EEAS}{\end{eqnarray*}}
\newcommand{\BEA}{\begin{eqnarray}}
\newcommand{\EEA}{\end{eqnarray}}
\newcommand{\BEQ}{\begin{equation}}
\newcommand{\EEQ}{\end{equation}}
\newcommand{\BIT}{\begin{itemize}}
\newcommand{\EIT}{\end{itemize}}

\title{Towards a Theory of Maximal Extractable Value II: Uncertainty}
\author{Tarun Chitra}
\date{September 2023}

\begin{document}

\maketitle
\begin{abstract}
Maximal Extractable Value (MEV) is value extractable by temporary monopoly power commonly found in decentralized systems.
This extraction stems from a lack of user privacy upon transaction submission and the ability of a monopolist validator to reorder, add, and/or censor transactions.
There are two main directions to reduce MEV: reduce the flexibility of the miner to reorder transactions by enforcing ordering rules and/or introduce a competitive market for the right to reorder, add, and/or censor transactions.
In this work, we unify these approaches via \emph{uncertainty principles}, akin to those found in harmonic analysis and physics.
This provides a quantitative trade-off between the freedom to reorder transactions and the complexity of an economic payoff to a user in a decentralized network.
This trade off is analogous to the Nyquist-Shannon sampling theorem and demonstrates that sequencing rules in blockchains need to be application specific.
Our results suggest that neither so-called fair ordering techniques nor economic mechanisms can individually mitigate MEV for arbitrary payoff functions.
\end{abstract}

\section{Introduction}
Blockchains and other decentralized value transfer systems involve coordinating networks of agents with varied reward and utility functions.
These systems inherently need to be designed to handle both sophisticated users, who optimize their transaction processing, and unsophisticated users who are not aware of how the precise microstructure of their transactions are processed.
While this problem exists in both traditional finance and online ad markets, blockchains have a unique problem as transaction submission is open to arbitrary users and all transaction data is eventually made public.
This transparency, while a boon for analysis of user behavior and fund dispersion, also creates excess profit opportunities for strategic users at the expense of non-strategic users.

However, such profits are not risk-free and are often needed for security properties in blockchain systems.
Arbitrage against decentralized exchanges is necessary to ensure that prices are kept in line with off-chain marketplaces~\cite{angeris2020improved}.
Strategic users generating MEV is crucial for the functioning of such arbitrage.
In some scenarios, it has even been demonstrated that there is a net positive externality to all users when such MEV is present as it improves execution and routing quality~\cite{kulkarni2022towards}.
On the other hand, MEV has been utilized to execute economic-based attacks on blockchain protocols and to reduce economic surplus to unsophiscated users.
There have been a number of articles that have attempted to quantify this~\cite{Builder, daian2020flash} and the conclusions on the magnitude of user impact are inconsistent.

\paragraph{MEV Mitigation.}
This lack of clear data on the net negative effect of MEV hasn't prevented research on MEV mitigation techniques.
One of the first (and by far, the least successful in practice) forms of MEV mitigation was so-called `fair ordering'~\cite{aequitas, kelkar2021themis}.
Fair ordering methods attempt to modify blockchain consensus to force validators to provide extra information about the relative times that they received particular transactions.
For instance, if more than 50\% of validators attest to receiving transaction A before transaction B, a fair ordering protocol could enforce (as a consensus rule) that transaction A preceeds transaction B.

Unfortunately, classical social choice theory shows that such pairwise ordering methods cannot always work due to Condorcet Paradoxes (and more generally, Arrow's impossibility theorem).
In response to this, fair ordering protocols attempt to create mechanisms that minimize the probability of such paradoxical occurrences.
However, it has now been established that such mechanisms are vulnerable to denial-of-service attacks~\cite{vafadar2023condorcet} and also distort economic payoffs for end users (\eg~there is an excess cost to the user in terms of lost utility that is greater than the gain from `fairness'~\cite{wadhwa2023breaking, angeris2023spectral}).

On the other hand, there are scenarios where ordering restrictions that are less strict than fair ordering can provide positive user outcomes.
For instance,~\cite{xavier2023credible} demonstrates that particular sequencing rules (\eg~restrictions on the final output ordering) for automated market markers can increase social welfare for unsophiscated agents.
This observation suggests that restricting orderings based on user welfare functions can mitigate the negative welfare effects of MEV.
Note that this is in contrast to fair ordering, which is agnostic to economic payoffs that occur under different orderings.

One natural question is how to construct sets of orderings that improve social welfare given a set of utility functions.
Both~\cite{kulkarni2022towards} and~\cite{xavier2023credible} are able to measure the impact of reordering on automated market makers in part by taking advantage of the convexity of the associated payoff functions.
The goal of this paper is to provide generalizations of these sequencing rules to a larger class of payoff functions.
We will demonstrate via example in~\S\ref{sec:payoffs} that the impact of sequencing rules can be dramatically larger for non-smooth and non-monotone payoffs.
These examples demonstrate that there is a trade-off between the complexity of an ordering rule and some (coarse) notion of the complexity of the payoff function realized by users.
In particular, these examples illustrate that more complex ordering rules tend to inure higher penalties (measured in terms of worst case payoffs to users) for less complex payoff functions.

\paragraph{Uncertainty Principles.}
One way of demonstrating that two functions $f, g$ are unable to both simultaneously be `simple' is to prove an \emph{uncertainty principle}.
While uncertainty principles are often associated with quantum systems in physics, they are a more general phenomena for linear operators.
In particular, if $C(f)$ is a measure of complexity of $f$, an uncertainty principle will usually be of  the form
\[
C(f) \cdot C(Lf) \geq c
\]
for a constant $c > 0$ and for a linear transform $L$.
The Heisenberg uncertainty principle corresponds to $C(f) = \frac{\int x^2 |f(x)| dx}{\Vert f \Vert_2}$ and $L$ being the Fourier transform.
Such an inequality says that if a complexity measure is `small' for $f$ then it cannot be small for $Lf$ beyond some level of precision (specified by the constant $c$).
For the Fourier transform, such uncertainty principles effectively state that $f$ and its Fourier transform $\hat{f}$ cannot both be localized.


We aim to construct complexity measures $C$ that represent fairness.
These measures will connect sizes of particular subsets of permutations, $A \subset S_n$, where $S_n$ is the set of permutations on $n$ elements to measures of complexity of the payoff function $f$.
We view the sets $A \subset S_n$ as the outputs of sequencing rules (such as fair ordering or the greedy CFMM rule of~\cite{xavier2023credible}).
A large set means that the ordering rule is not restricting orderings sufficiently whereas a small set is highly constrained.
While there are many inequivalent notions of function complexity for functions $f : S_n \rightarrow \reals$, we will aim to choose the simplest measure based on the boolean degree of a function~\cite{filmus2020hypercontractivity}.

\paragraph{This Paper.}
Using uncertainty principles as our guide, we will first reduce the problem of measuring welfare loss due to MEV to a discrete harmonic analysis problem.
We do this by considering real-valued payoff functions that take in a sequence of transactions from a fixed set, $f(T_1, \ldots, T_n)$.
We then consider a functional that maps a function $f$ and a set of orderings $A\subset S_n$ to $[0, 1]$ to measure the `fairness' of restricting $f$ to $A$.
One can use this measurement to make relative comparisons of fairness for given payoff functions and orderings $A$.

Our main results are that the fairness functionals that we define in~\S\ref{sec:payoffs} can be upper and lower bounded using discrete harmonic analysis (such as uncertainty principles).
More precisely, we look at a payoff $f$ and its Fourier transform $\hat{f}$ and utilize existing uncertainty principles to show that the fairness functional of a payoff is bounded by the $L^1$ and $L^{\infty}$ norms of $\hat{f}$.
This allows us to precisely state when a set of orderings $A \subset S_n$ worsens the fairness functional of $f$.

In the process, we also construct two major examples of payoff functions that are commonly found in decentralized finance (DeFi): constant function market makers (CFMM) and liquidations.
The fairness and MEV properties CFMMs have been repeatedly studied (\eg~\cite{kulkarni2022towards, xavier2023credible, zhou2021high, chitra2022differential}) as their payoffs are smooth and have convexity properties.
Liquidations, on the other hand, are much harder to study because their payoff are extremely non-smooth.
We utilize recent work~\cite{angeris2023spectral} that demonstrates that liquidations can be used as function basis for the set of reordering MEV payoff functions $f$.

One natural interpretation is that the Fourier expansion of payoff function corresponds to expanding it in a basis of liquidations.
This allows us to have concrete realizations of payoff functions, represented as combinations of liquidations, that can saturate the best and worst case outcomes measured by fairness functionals.
We believe that this `liquidation representation' of an MEV payoff (which is just its Fourier transform over the symmetric group) will be useful for further design and analysis of MEV.

\paragraph{Notation.}
The set of permutations of $n$ elements will be denoted $S_n$.
For a permutation $\pi \in S_n$, $\pi(i)$ refers to the index where the $i$th element is moved.
The probability simplex will be defined as $\Delta^n = \{(x_1, \ldots, x_n): x_i \geq 0, \sum_i x_i = 1\}$.
We denote by $[n] = \{1, \ldots, n\}$ and $\lambda = (\lambda_1, \ldots, \lambda_k) \vdash n$ a partition of $n$, \eg~$\sum_i \lambda_i = n$ with $\lambda_i \in \naturals$.
For a finite set $S$, we denote the expectation of a function $f : S \rightarrow \reals$ as $\Expect[f] = \frac{1}{|S|} \sum_{s \in S} f(s)$.
We treat $\reals^{n!}$ as the set of functions $f: S_n \rightarrow \reals$ and view $L^p(S_n)$ as $\reals^{n!}$ with the metric $\Vert \cdot\Vert_p$.

\section{Background}\label{sec:background}
The goal of this section is to describe how to reduce a large portion of MEV to a problem about functions $f: S_n \rightarrow \reals$.
Reducing MEV to a concrete mathematical problem allows us to use tools from algebra and combinatorics to provide concrete guarantees.
We note that our definitions only focus on reordering MEV as opposed to MEV arising from censorship of transactions.
However, our algebraic tools can be extended to handle censorship with some extra care (this is addressed in~\S\ref{conclusion}).

\subsection{Blockchains.}
Blockchains are decentralized systems that come to consensus on particular state and execution of computation on that state.
We assume that readers are familiar with the basic notions of blockchains and provide the minimal description.
Readers interested in more detials are referred to modern textbooks such as~\cite{shifoundations} for further details.

A blockchain consists of a monotonically increasing sequence of blocks $B_i$, $i \in \naturals$ that contain state and state transitions.
Each block is made up of transactions $T_1, \ldots, T_n$ that mutate the state of the previous block and/or write new state.
Users submit their transactions via a peer-to-peer network and pay a fee to have their transactions included in a block.
Validators or miners are network participants who collect transactions and come to consensus about the validity of a block $B_i$.
Validators or miners must lock a resource to participate and are incentivized with fees from user transactions and from a subsidy known as a block reward.
One key aspect of blockchains is that they allow for asynchronous communication about shared state and that they are designed such that it is costly for validators to deviate from the consensus protocol.

Different consensus mechanisms make different assumptions about validators in order to guarantee security properties of the blockchain.
For instance, an honest majority assumption is necessary to prove that Proof of Work (which is what Bitcoin uses) provides users with safety and liveness.
Safety is the property that once a transaction has entered a confirmed block, it cannot be evicted later (with high probability).
Liveness is the property that the network can continually take new transactions and that the time from when a user submits a transaction to when it is confirmed is bounded.

One common feature to many consensus protocols is that a single validator holds a monopoly over the production of block $B_i$.
A randomness beacon is used to select a random validator $V_i \sim S \in \Delta^{V}$ where $V \in \naturals$ is the number of validators and $S$ is a distribution over locked resources.
Once a validator $V_i$ is chosen, they have the right to choose which transactions $T_1, \ldots, T_n \in \mathcal{T}$ can be included and the order in which those transactions are included.
Most consensus mechanisms, such as Proof of Work or Proof of Stake, have the probability of a validator $V_i$ being chosen proportional to the amount of resource they locked.
This way, validators who are contributing more to the network's security are rewarded with a higher pro-rata portion of network fees and inflation.

\subsection{MEV}
The temporary monopoly guaranteed to a single validator affords them the opportunity to include, exclude, or reorder transactions to maximize a validator's profit.
Any strategy that deviates from ordering transactions by the fees they pay can be viewed as a \emph{maximal extractable value} strategy.
As noted in~\cite{bahrani2023transaction}, the precise definition of MEV can be somewhat subtle to describe.
However, for this paper we will view any strategy that does not order transactions based on transaction fees and earns a higher expected profit for validators as an MEV strategy.

The types of strategies that can be employed by validators differs depending on the types of transactions $T_i$ that are submitted.
Some strategies involve adding front-running and `back-running' transactions around user trades in order to allow arbitrage trades to take advantage of uninformed user flow.
This type of MEV has been well-studied and in fact bounds on social welfare change were calculated in~\cite{kulkarni2022towards}.

In this paper, we will focus on MEV that arises from reordering.
We assume that there is a fixed set of transactions $T_1, \ldots, T_n$ and a payoff function $f(T_1, \ldots, T_n)$ that yields the payoff to a validator for a particular ordering.
We note that we consider the entire amount of MEV extracted and do not consider the allocation of MEV to validators versus to `searchers', who are agents who submit orderings to validators and are paid as a function of the excess profit they generate.

\subsection{Payoff Functions.}\label{sec:payoffs}
Since we consider a fixed universe of transactions in reordering MEV, we can view the payoff function for the validator as a function $f : S_n \rightarrow \reals$.
Concretely, if we have a function $g : \mathcal{T}^n \rightarrow \reals$ that maps transactions to payoffs, we can define the function $f$ as
\[
f(\pi) = g(T_{\pi(1)}, \ldots, T_{\pi(n)})
\]
One can view the the ordering $\max_{\pi \in S_n} f(\pi)$ as the optimal monopoly profit achievable via reordering MEV.
We will assume that all functions $f : S_n \rightarrow \reals$ are positive, $f(\pi) \geq 0$, noting that our analysis applies to functions that are bounded below by a constant, $f(\pi) \geq -c,\,\forall \pi \in S_n$.
Before continuing, we will first provide two examples of payoff functions.

\paragraph{Example: Constant function market makers.}
Suppose that there is a decentralized exchange, such as a constant function market maker (CFMM)~\cite{angeris2020improved}, that users can tender assets to for trading.
A sandwich attack~\cite{zhou2021high, kulkarni2022towards} is a type of transaction where a validator inserts a transaction before a user trade to increase the price ahead of a user trade. After the execution of the user trade, the validator inserts a trade going in the opposite direction to generate a profit.
This type of front running attack involves three types of transactions: $F_i, \Delta_i, B_i$, where $F_i$ is the front-running transaction, $B_i$ is the back-running transaction, and $\Delta_i$ is the unsophisticated user trade.
Owing to the concavity properties of a CFMM, one can upper and lower bound the profit of a single sandwich attack via a linear function of $\Delta_i$ and the current price $p_i$.

Note that if a validator receives $n$ trades $\Delta_1, \ldots, \Delta_n$ then the order in which they execute the sandwich attacks has material impact on their profit.
For instance, executing larger size trade earlier makes the profitability of smaller trades executed later in the ordering higher.
This allows one to write a payoff function of the form
\[
f(\pi) = \sum_{i=1}^n PNL(\Delta_{\pi(i)}, p_{\pi(i-1)})
\]
where $p_i$ is the price tendered after executing $\Delta_i$ and we abuse notation slightly and interpret $p_{\pi(0)}$ as the initial price from the previous block.
This payoff is generally not invariant to permutations as the price series $p_i$ is dependent on the previous set of trades $(\Delta_1, \ldots, \Delta_{i-1})$ and each trade causes positive price impact.

\paragraph{Example: Liquidations.}
Liquidations are transactions used to clear overindebted positions in collateralized lending and perpetual protocols within DeFi.
Briefly, a user posts $q_c$ units collateral of asset X and borrows $q_b$ units of asset Y when the price of X in units of Y is $p_0$.
Suppose the price of X in units of Y at time $t$ is $p_t$.
If the price decays such that $\nicefrac{q_c}{p_t} < q_b$ (\eg~the collateral is worth less than the borrowed asset), then a liquidator can submit a transaction to the blockchain in which they supply $(1-\epsilon)q_b$ units of Y and get $q_c$ units of X back.
We call the price threshold $p^* = \nicefrac{q_c}{q_b}$ the liquidation price of the position.

Consider a scenario where there are $2k$ trades $\Delta_i$ with $k$ trades causing the price to go up by 1 unit and $k$ trades causing the price to go down by 1 unit.
Suppose that the initial price prior to executing trades $\Delta_i$ is $p_0$ and that $p^* = p_0 - c$ where $c < k$.
Furthermore, given an ordering $\pi \in S_n$ define the price $p_i(\pi)$ to be the price after executing $\Delta_{\pi(1)}, \ldots, \Delta_{\pi(i)}$.
Note that for some permutations $\pi$, the liquidation is not possible (\eg~permutations where all of the trades that move the price up are executed before those that go down) and for others it is possible.
Define the liquidatable set $A = \{ \pi \in S_n : \exists i \; p_i(\pi) \leq p_0 -c\}$.
Then our payoff function is
\[
f(\pi) = \ones_A(\pi)
\]
More generically, as described in~\cite{angeris2023spectral}, one can utilize liquidations and auctions to construct payoffs of the form $\ones_B$ for any $B \subset S_n$.
Since any function $f(\pi)$ can be written as
\[
f(\pi) = \sum_{A \subset S_n} \hat{f}(A) \ones_A(\pi)
\]
for some coefficients $\hat{f}(A) \in \reals$, this implies that liquidations can be thought of as generating a basis for the set of functions $f : S_n \rightarrow \reals$.
We will later see that the Fourier transform directly expands $f$ in a basis like this, allowing us to interpret the Fourier coefficients as a basis expansion in the `liquidation' basis.

\paragraph{Fairness Functionals.}
Given a payoff function $f : S_n \rightarrow \reals$, one natural question to ask is how to measure the fairness of an ordering scheme that generates a set $A \subset S_n$ of valid orderings.
To do this, we need to measure some notion of worst and best case payoffs over the set $A$.
Towards this aim, we first define a \emph{global fairness functional} as a map $\Lambda : L^1(S_n)\rightarrow \reals$ which takes a payoff for a set of transactions, and returns a deviation between the maximum and average case behavior.

The two most natural global fairness functionals are:
\begin{align*}
    \tilde{\Lambda}^+(f) &= \max_{\pi \in S_n} f(\pi) - \frac{1}{n!}\sum_{\pi \in S_n} f(\pi) = \Vert f \Vert_{\infty} - \frac{1}{n!}\Vert f \Vert_1 = \Vert f \Vert_{\infty}  - \Expect[f]\\
    \tilde{\Lambda}^{\star}(f) &= \frac{\max_{\pi \in S_n} f(\pi)}{\frac{1}{n!}\sum_{\pi \in S_n} f(\pi)} = \frac{\Vert f \Vert_{\infty}}{\frac{1}{n!}\Vert f \Vert_1} = \frac{\Vert f \Vert_{\infty} }{\Expect[f]}
\end{align*}
The first functional, $\tilde{\Lambda}^+$ measures the additive difference between the maximum value of the function $f$ and expected value whereas the latter measures the multiplicative gap.
The latter is similar to a price of anarchy measurement; however, it lacks Lipschitz properties and isn't directly related to the graph Laplacian for a Cayley graph of $S_n$.
On the other hand, the authors of~\cite{angeris2023spectral} demonstrate that these properties hold for $\tilde{\Lambda}^+$.

We can interpret a function that has $\Lambda^+(f) = 0$ as perfectly fair in that returning a random ordering has the same payoff as the best case payoff.
On the other hand, note that $\Expect[f] \geq \frac{1}{n!}\max_{\pi \in S_n} f(\pi)$ so $\tilde{\Lambda}^+(f) \leq \max_{\pi \in A}f(\pi) \left(1- \frac{1}{n!}\right)$.
A function $f$ has \emph{maximal unfairness} if it saturates this bound, which happens if $f = \ones_{\pi}$ for some $\pi \in S_n$.
We say that a payoff is asymptotically maximally unfair if $\max_{\pi \in A}f(\pi) \left(1- \frac{1}{n!}\right) - \Lambda^+(f, A) = \Theta\left(\frac{1}{n!}\right)$
We call perfectly fair functions and asymptotically maximally unfair functions \emph{asymptotically trivial payoffs} for $A$.

We define localized fairness functionals $\Lambda : L^1(S_n) \times 2^{S_n} \rightarrow \reals$ that are localized to a set $A \subset S_n$ as
\begin{align*}
\Lambda^+(f, A) &= \tilde{\Lambda}^+(f \ones_A) = \max_{\pi \in A} f(\pi) - \Expect[f\ones_A] \\
\Lambda^{\star}(f, A) &= \tilde{\Lambda}^{\star}(f \ones_A) = \frac{\max_{\pi\in A} f\ones_A(\pi)}{\Expect[f\ones_A]}
\end{align*}
We note that prior work on social welfare guarantees for CFMMs~\cite{kulkarni2022towards} studied $\tilde{\Lambda}^*(f)$ and showed that provided sufficient liquidity, $\tilde{\Lambda}^*(f) = \Theta(\log n)$.

The remainder of the paper will be dedicated to showing asymptotically non-trivial upper and lower bounds on $\Lambda^+(f)$.
The lower bounds represent a minimal amount of unfairness injected by the selection of an ordering set $A \subset S_n$ whereas the maximal bounds represent a non-trivial fairness guarantees.
Our results in~\S\ref{sec:results} will show that if $|A|$ is sufficiently `large', then one can achieve non-trivial fairness guarantees whereas if $|A|$ is too `small' then one has fairness lower bounds for $\Lambda^+$.
To define the notion of what a `large' and `small' mean, we will need to using representation theory to define the concept of the \emph{boolean degree} of a function.

We finally note that the bounds we construct actually bound $\Lambda^*(f, A)$, which is then used to bound $\Lambda^+(f, A)$ since
\begin{align}\label{eq:connecting-bds}
\Lambda^+(f, A) &= \max_{\pi \in A} f(\pi) - \Expect[f\ones_A] = \max_{\pi \in A} f(\pi) \left(1 - \frac{\Expect[f \ones_A]}{\max_{\pi \in A}f(\pi)}\right) \nonumber \\
&= \max_{\pi \in A} f(\pi) \left(1 - \frac{1}{\Lambda^{\star}(f, A)}\right)
\end{align}

\subsection{Representation Theory and Uncertainty Principles}
If one is given black-box access to a payoff function $f : S_n \rightarrow \reals$, how can one figure out if the maximum and expected values deviate from one another?
The liquidation example of~\S\ref{sec:payoffs} noted that any payoff function $f : S_n \rightarrow \reals$ can be expanded as a sum of indicator function.
Our goal is to see if there is a small set of permutations that controls the behavior of the payoff $f : S_n \rightarrow \reals$.
For instance, if there is a set $A \subset S_n$ with $|A| = O(1)$ such that $f(\pi) = \ones_A(\pi)$, then there is a large separation between the optimal payoff of $1$ and the average payoff of $\frac{|A|}{n!} = O(\frac{1}{n!})$.
This large separation between $\max_{\pi \in S_n} f(\pi)$ and $\Expect_{\pi \in S_n}[f(\pi)]$ is precisely what fairness functionals aim to measure and is completely controlled by $|A|$ for indicator function payoffs.

Suppose that one could decompose a general payoff $f : S_n \rightarrow \reals$ into a sum of indicator functions on sets $A \subset S_n$, $f = \sum_{A\subset S_n} \hat{f}(A) \ones_A$ and quantify the size of the coefficients $|\hat{f}(A)|$.
This would allow one to take a set of permutations, possibly constructed via a sequencing rule or auction, and measure the fairness as illustrated by the previous example.
The Fourier Transform over finite groups precisely characterizes how to decompose a function into a basis of indicator functions and compute these coefficients.
In particular, this allows for one to provide quantitative control over how a particular set of orderings $A$ impacts a fairness functional as measured by the sizes of set $|A|$ versus the size of their coefficients.

\paragraph{Fourier-Walsh Transform and boolean degree.}
If $f$ were a boolean function, $f : \{-1, 1\}^n \rightarrow \reals$, then the expansion $f(x) = \sum_{A \subset [n]} \hat{f}(A) \ones_A(x)$ where $\ones_A(x)$ is one is $x_i = 1$ for all $i\in A$ is known as the Fourier-Walsh transform~\cite{o2014analysis}.
One can view this as a discrete analogue of the classical Fourier transform for discrete spaces.
We will first describe some properties of the Fourier-Walsh transform over boolean functions $\{-1, 1\}^n$ before moving to functions $f : S_n \rightarrow \reals$.
The boolean case is easier to understand and will provide intuition for what we can expect in the symmetric group case.

Using the Fourier-Walsh transform, one can prove statements that tie the global behavior of the function $f$ to properties about the sets $A$ that it is supported on.
For instance, if $f$ is a voting rule where each $x_i$ is a vote for one of two candidates and the output is an aggregate vote (such as majority or weighted majority vote), then the sets $A \subset \{-1, 1\}^n$ with large values of $\hat{f}(A)$ control the outcome.
Classical results in voting theory such as Arrow's impossibility theorem can be made quantitative via looking at them via their Fourier-Walsh transform.
If $i\in A$ for all $A$ with $\hat{f}(A) \neq 0$, then we say that $i$ is a dictator.
One can show that if $f$ is monotone and unanimous, there exists $i$ such that $i \in A$ for any $A$ with $\hat{f}(A) \neq 0$ (\ie~there is a dictator, see~\cite[\S2.5]{o2014analysis} for a precise statement). 

Note that the Fourier-Walsh transform inversely relates the size of a set $A$ to its influence on $f$.
To see this, suppose $A \subset [n]$ is such that $|A| = 2$.
Then there are $2^{n-2}$ vectors $v$ in $\{-1, 1\}^n$ such that $\ones_{A}(v) = 1$.
One can generalize this and show that the logarithm of the support of the function $\ones_A$ is equal to $n - |A|$.
Thus if there is a small set $|A|$ that has a large Fourier coefficient $\hat{f}(A)$, then a small subset of elements influence the outcome of $\hat{f}(A)$.

We can thus interpret the relationship between $|A|$ and the magnitude of $\hat{f}(A)$ as representing how `flat' or `sharp' the function $f$ is.
Moreover, we have the Plancharel theorem for the Fourier-Walsh transform, just as we do for real-valued functions:
\[
\sum_{x\in\{-1,1\}^n} |f(x)|^2 = \sum_{A\subset [n]} |\hat{f}(A)|^2
\]
This means that if some small collection of sets $A \subset [n]$ has more than $(1-\epsilon)$ of the norm of $f$, then the function is `essentially' controlled by a small group of variables.

For such an expansion, we define the boolean degree, $\deg(f)$, as
\[
\deg(f) = \max\{ |A| : A \subset \{-1, 1\}^n, |\hat{f}(A)| > 0\}
\]
Intuitively, the degree of a function measures the size of the largest subset that controls function behavior.
We can also restrict a function to its low degree set.
For any $t \in [n]$, we define the degree-$t$ restriction $f^{\leq t}(x)$ as
\[
f^{\leq t}(x) = \sum_{\substack{A \subset \{-1, 1\}^n \\ |A| \leq t}} \hat{f}(A) \ones_A(x)
\]
and define the complement $f^{> t} = f - f^{\leq t}$.
Note that by definition, $\deg(f^{\leq t}) \leq t$.
Our notions of `large' and `small' sets for the bounds in~\S\ref{sec:results} will be defined based on $\deg(f)$ 

\paragraph{Fourier Analysis of the Symmetric Group}
A similar expansion exists for the symmetric group $S_n$, although it is significantly more complex and involved.
Since the symmetric group is non-abelian --- $\pi \sigma \neq \sigma \pi$ for two permutations $\pi, \sigma \in S_n$ --- indicator functions become matrices.
Instead of having indictator functions $\ones_A : \{-1, 1\}^n \rightarrow \reals$, one has representations $\rho : S_n \rightarrow \mathsf{GL}(V^d)$, where $\mathsf{GL}(V^d)$ is the set of invertible matrices on a vector space of dimension $d$.
For brevity, we will ignore most of the algebraic properties of representation theory and present the bare minimum to understand the proofs of~\S\ref{sec:results}.
The interested reader can see the textbooks~\cite{diaconis1983mathematics, sagan2013symmetric} for a full treatment.

A representation of a finite group $G$ is a map $\rho : G \rightarrow \mathsf{GL}(V^d)$, where $d$ is the dimension of a representation.
Two representations $\rho_1 : G \rightarrow \mathsf{GL}(V^{d_1})$, $\rho_2 : G \rightarrow \mathsf{GL}(V^{d_2})$ can be summed as $\rho = \rho_1 \oplus \rho_2$ which is a matrix in $\mathsf{GL}(V^{d_1 + d_2})$.
A representation $\rho$ is irreducible if there exist no representations $\rho_1, \rho_2$ such that $\rho = \rho_1 \oplus \rho_2$.
It is a fact that finite groups have a finite number irreducible representations $\rho_1, \ldots, \rho_k$ and they always satisfy the formula
\[
\sum_{i=1}^k \dim(\rho_i)^2 = |G|
\]
Representations will serve as the analogues of indicator functions\footnote{The indicator functions in the boolean case are a special case of \emph{characters} of a finite abelian group} for non-abelian groups.

Given a set of irreducible representations $\rho_1, \ldots, \rho_k$ for $S_n$ one defines the Fourier transform $\hat{f}$ of a function $f : S_n \rightarrow \reals$ to be:
\begin{align*}
    \hat{f}(\rho) = \sum_{\pi \in S_n} f(\pi) \rho(\pi)
\end{align*}
Note that this is a sum of matrices and hence $\hat{f}(\rho) \in \mathsf{GL}(V^{\dim(\rho)})$.
Given the Fourier transform $\hat{f}(\rho)$ one can invert the function $f$ via the inversion formula~\cite{diaconis1988group},
\begin{equation*}
f(\pi) = \frac{1}{n!}\sum_{i=1}^k \dim(\rho_i) \Ttr[\hat{f}(\rho_i) \rho_i(\pi)] 
\end{equation*}
Similarly to the boolean case, we have a version of the Plancherel theorem,
\[
\Vert f\Vert_2 = \frac{1}{n!}\sum_{i=1}^k \dim(\rho_i)\Ttr[\hat{f}(\rho_i)^* \hat{f}(\rho_i)]
\]

In order to define an analogue of the boolean degree, we will need to have a more concrete definition of the irreducible representations of $S_n$.
Note that the set of partitions of $[n]$, $\lambda_1 \geq \lambda_2 \geq \cdots \geq \lambda_k$ such that $\sum_i \lambda_i = n$, describes the set of irreducible representations of $S_n$.
This is because every permutation can be decomposed into a product of cycles $C_1 \cdots C_k$ that partition $n$ and such that elements transposed by $C_i$ are not ever transposed with $C_j$ for $i \neq j$.
To signify that a vector $\lambda$ is a partition of $[n]$, we use the standard notation $\lambda \vdash n$.
We can write expand the function via the inverse Fourier transform as
\[
f(\pi) = \frac{1}{n!} \sum_{\lambda \vdash n} d_{\lambda} \Ttr[\hat{f}(\rho^{\lambda}) \rho^{\lambda}(\pi)] = \frac{1}{n!} \sum_{\lambda \vdash n} d_{\lambda} f^{=\lambda}(\pi) 
\]

It turns out that each partition $\lambda \vdash n$ defines a set of functions invariant under permutations whose cycle decomposition is $\lambda$.
This set, called the Specht module~\cite{sagan2013symmetric} $S^{\lambda}$, allows one to decompose $L^1(S_n) = \bigoplus_{\lambda \vdash n} S^{\lambda}$.
Thus the sum above represents the projection of $f$ to each Specht module.
We are now in a position to define the boolean degree of a function $f$~\cite{filmus2020hypercontractivity}:
\begin{equation}
\deg(f) = \min\{ n-\lambda_1 : \lambda \vdash n, |f^{=\lambda}| > 0\}
\end{equation}
Intuitively, this degree measures the total entropy of the function outside of the largest component and we refer the interested reader to~\cite{filmus2020hypercontractivity} for more details on how why this is the correct analogue of boolean degree for the symmetric group.

As a concrete example of functions of a particular degree, we introduce the $k$-juntas~\cite{filmus2020hypercontractivity}.
The set of $k$-juntas $J_{k,n}$ are linear combinations of products of indicator functions on $k$ unique transpositions,~\ie
\[
J_{k,n} = \left\{ \sum_{T = \{(i_1,j_1) \cdots (i_k, j_k)\} \in \binom{n}{k}^2} a_{T} \prod_{\ell=1}^k \ones_{\pi(i_{\ell}) = j_{\ell}} \right\}
\]
The $k$-juntas have degree $k$ and span the set of all degree-$k$ functions.

Our results in \S\ref{sec:results} will compare the size of $A \subset S_n$ to $\deg(f)$.
If a particular monotone function of $|A|$ is smaller than $\deg(f)$, then we will show fairness functional lower bounds.
We will show the opposite if the same function of $|A|$ is greater than $\deg(f)$.

\paragraph{Uncertainty Principles.}
One important property that relates functions and their Fourier transforms are \emph{uncertainty principles}.
Given a function $f$ and a Fourier transform $\hat{f}$, these principles state that both $f$ and $\hat{f}$ cannot be `localized' simulataneously beyond a point.
For instance, the classical Heisenberg uncertainty principle says that $\Var(f)\Var(\hat{f}) \geq c$ for a real function $f: \reals \rightarrow \reals$.
One can also view these as saying that if $f$ is concentrated on a small set $S$ of values then $\hat{f}$ cannot be concentrated on a set of values more than some decreasing function of $\mu(S)$ for Lebesgue measure $\mu$.

There exist uncertainty principles for finite groups as well and the following recent result of Kuperberg (first described in~\cite{wigderson2021uncertainty}) provides a direct connection to the $L^1$ and $L^{\infty}$ norms of a function $f$ and its Fourier transform $\hat{f}$:

\begin{theorem}[Kuperberg, Wigderson, Wigderson~\cite{wigderson2021uncertainty}]\label{thm:kuperberg}
Let $G$ be a finite group and consider $f : G \rightarrow \reals$. Let $\rho_i : G \rightarrow V_i$ be a complete set of (complex) irreducible representations for $G$. Then we have
\begin{equation}\label{eq:uncertainty}
\frac{\Vert f \Vert_1}{\Vert f \Vert_{\infty}} \frac{\Vert \hat{f} \Vert_1^{(S)}}{\Vert \hat{f} \Vert_{\infty}^{(S)}} \geq |G|
\end{equation}
where $\Vert A \Vert_p^{(S)} = \Ttr[(A^*A)^{p/2}]^{1/p}$ is the Schatten $p$-norm of a matrix $A$
\end{theorem}
\noindent If $G = S_n$, then this states that
\[
\frac{\Vert f \Vert_1}{\Vert f \Vert_{\infty}} \frac{\Vert \hat{f} \Vert_1^{(S)}}{\Vert \hat{f} \Vert_{\infty}^{(S)}} \geq n! \Longrightarrow \frac{1}{\tilde{\Lambda}^{\star}(f)} = \frac{\frac{1}{n!}\Vert f \Vert_1}{\Vert f \Vert_{\infty}} \geq \frac{\Vert \hat{f} \Vert_{\infty}^{(S)}}{\Vert \hat{f} \Vert_{1}^{(S)}}
\]
Therefore, uncertainty principles immediately furnish upper bounds on $\tilde{\Lambda}^{\star}$ which correspond to upper bounds on $\tilde{\Lambda}^+$ via \eqref{eq:connecting-bds}.

\subsection{$t$-intersecting sets of permutations}\label{sec:t-intersecting}
In order to formally define the notions of `small' and `large' sets $|A|$ for a payoff, we need to consider the amount of overlap between elements of $A$.
One can informally think of a measure overlap of elements of $A$ as the number of points $i$ that are moved to a point $j$ by permutations in $A$.
Formally, a set $A \subset S_n$ is \emph{$t$-intersecting} if for all $\pi, \pi' \in A$, there exist $t$-pairs $(i_1, j_1), \ldots, (i_t, j_t)$ such that $\pi(i_k) = \pi'(i_k) = j_k$.
Note that the sets of pairs of elements of $[n]$ can vary for different pairs of permutations $\pi, \pi' \in A$.

It was recently shown that sufficiently large $t$-intersecting sets of permutations $A$ have their size bounded by $(n-t)!$~\cite[Thm. 1]{keller2023t}
We claim that this result implies that indicator functions $\ones_A$ for such sets $A$ have degree that is always larger than $t$~\cite[Thm. 1]{keller2023t}.
This means that we can view restrictions of functions to $t$-intersecting sets as a form of `bandlimiting' in that it restricts the cycle types of permutations that can have a large Fourier coefficient.
We formalize this claim as follows:
\begin{claim}\label{claim:indicator-degree}
Suppose $A \subset S_n$ is a $t$-intersecting set of permutations such that $c_0 t \leq n$ where $c_0$ is the universal constant of~\cite[Thm. 1]{keller2023t}.
If $|A| = \Omega((n-t)!)$ then $\deg(\ones_A) \geq t$. 
\end{claim}
The proof is a simple corollary of~\cite[Thm. 1]{keller2023t} but is very important for our proofs in~\S\ref{sec:results}.
In particular, this ensures that if $\deg(f) \geq \deg(\ones_A)$ then there are Fourier terms summing over partitions $\lambda \vdash n$ with $\lambda_1 \in [n-t, n-\deg(f)]$ that can potentially have very high unfairness (as measured via $\Lambda^+$).

\paragraph{Relationship to Fair Ordering.}
These sets of permutations are natural in that they often occur within ordering algorithms and MEV.
Fair ordering protocols~\cite{kelkar2021themis, ramseyer2023fair} work by having validators come to consensus on a particular directed graph $G$.
Provided there is at least an honest majority, the validators of the network can take any valid topological sort of $G$ and return it as a valid ordering.
As noted by~\cite{ramseyer2023fair}, the graphs constructed are relatively simple.
The graph $G$ has a vertex set equal to the set of transactions, $V = \{T_1, \ldots, T_n\}$ and weighted directed edges $(T_i, T_j)$ with weight equal to the number of validators who say that $T_i$ arrived before $T_j$.

However, such topological sorts are still subject to the standard impossibility theorems of social choice theory.
For instance, the Condorcet paradox can still arise in such orderings, where the set of topological sorts contains $T_i > T_j, T_j > T_k, T_k > T_i$.
This leads to the existence of cycles that are conserved within the set $A$ of valid topological sorts of $G$.
In particular, there exists a minimal $t \in [n]$ such that set of sorts is $t$-intersecting with $t < n$ when there exist Condorcet cycles.

It is natural that there the set of topological sorts is $t$-intersecting in that there are likely transactions that all validators received before other transactions.
In fact, the set of topological sorts is only $0$-intersecting if there are no edges or if all edges have exactly 50/50 splits on validators' votes on transaction order precedence.
As such, we can view the results of~\S\ref{sec:results} as placing bounds on the `fairness' of fair ordering (as measured by fairness functionals) for particular payoff functions.


\section{Main Results}\label{sec:results}
We have two results on fairness: an upper bound on unfairness if the set $A$ is $t$-intersecting and $t$ is more than the degree of $f$ and a lower bound otherwise.
One can view these as analogues of Nyquist-Shannon sampling theorems in that one needs to have a sufficiently small set of permutations to resolve high degree modes.

\paragraph{Upper Bound.}
Our upper bound result shows that if the set of permutations has high pairwise overlap, then one gets an upper bound that is asymptotically less than the trivial upper bound of $\left(1-\frac{1}{n!}\right) \Vert f \Vert_{\infty}$.
\begin{claim}\label{claim1}
Suppose that $A \subset S_n$ is a $t$-intersecting set of valid orderings and $f : S_n \rightarrow \reals$ is an admissible payoff function.
Then if $t \geq \deg(f)$ there exists constants $C > 0, c > 1$ such that
\begin{equation}\label{eq:claim1}
\Lambda^+(f, A) = \Vert f \ones_A \Vert_{\infty} - \frac{1}{n!}\Vert f \ones_{A} \Vert_1 \leq \left(1 - \frac{C}{\binom{n}{s}^2 c^{\sqrt{s}}  s!}\right) \Vert f \ones_A \Vert_{\infty}
\end{equation}
where $s = \deg(f)$
\end{claim}
We note that this is an asymptotically non-trivial upper bound, as $\binom{n}{s}^2 c^{\sqrt{s}} s! = O(n^{2s} c^{\sqrt{s}} s!) = o(n!)$ if $s = o(n)$.
The proof of claim~\ref{claim1} relies on the uncertainty principle for finite groups, Theorem~\ref{thm:kuperberg}.

We will sketch how this theorem provide us an upper bound here but leave the full proof to Appendix~\ref{proof-claim-1}.
Let $k(n, s) = \frac{\Vert \hat{f} \Vert_{\infty}^{(S)}}{\Vert \hat{f} \Vert_{1}^{(S)}}$ and note that the $t$ dependence will drop out due to $t \geq s$.
Then Equation~\eqref{eq:uncertainty} implies that
\[
\frac{\Vert f \Vert_1}{\Vert f \Vert_{\infty}} \geq |G| k(n, s) = n! \cdot k(n, s)
\]
Dividing by $n!$ yields
\[
\frac{\Expect[f]}{\Vert f \Vert_{\infty}} \geq k(n, s)
\]
which implies that $-\Expect[f] \leq -k(n,s) \Vert f \Vert_{\infty}$ so that we have
\[
\Vert f \Vert_{\infty} - \Expect[f] \leq (1-k(n, s)) \Vert f \Vert_{\infty}
\]
The remainder of the proof (see Appendix~\ref{proof-claim-1} involves using facts from representation theory and the fact that $f = f^{\leq s}$ and $t \geq s$ to show that $k(n, s)$ has the form of~\eqref{eq:claim1}.

\paragraph{Lower Bound.}
On the other hand, the lower bound states that if the set of permutations does not have sufficient overlap (relative to the degree of $f$), then there is always a large amount of unfairness (as measured by $\Lambda^+$).
\begin{claim}\label{claim:claim2}
Let $A$ be a $t$-intersecting set $A \subset S_n$ with $|A| = \Omega((n-t)!)$ and $f: S_n \rightarrow \reals$ be a payoff. If $t < \deg(f)$, there exists $c, c' > 0, l \in [t, s]$ such that
\begin{equation}
\Lambda^+(f, A) = \Vert f \ones_A \Vert_{\infty} - \frac{1}{n!}\Vert f \ones_{A} \Vert_1 \geq \left(1-\frac{c'(s-t-1)}{(n-t)!}\right) \Vert f \ones_A \Vert_{\infty}
\end{equation}
\end{claim}

Note that the assumption of $|A| = \Omega((n-t)!)$ is non-trivial.
It is clear than one can have arbitrarily small $t$-intersecting sets (\eg~size 2), but we note that if $|A| = \Omega((n-t)!)$ then~\cite{keller2023t} demonstrated that there exist $(i_1, j_1), \ldots, (i_t, j_t)$ such that for all $\pi \in A$, $\pi(i_k) = j_k$.
This has implications for fair ordering in that fair ordering protocols attempt to have agreement on a maximal set of elements (\ie~want the $t$-intersecting property to hold for $t = \Omega(n)$) yet also need sufficiently large sets $A$ to achieve agreement.

The proof of Claim~\ref{claim:claim2} utilizes the fact that the partitions of $[n]$ also furnish a decomposition of the function space $L^1(S_n)$ into a series of subspaces $V_i$.
We project the payoff onto each subspace $V_i$ and then use bounds on the eigenvalues of a particular random walk on $V_i$, inspired by similar techniques used in~\cite{filmus2020hypercontractivity}.
We note that the representation theory utilized is relatively minimal and an open question is whether one can prove (or improve) these bounds via purely combinatorial means.

\paragraph{Examples.}
As a concrete example of the bound, consider $f = \ones_B$ with $B$ $t$-intersecting and $|B| = \Omega(n-t)!)$ and $A$ such that $|A \cap B| = 1$.
If $A$ is $t$-intesecting and $|A| = \Omega((n-t)!)$ then $\deg \ones_B = \deg \ones_A$ and the upper bound satisfies
\[
\Lambda^+(f, A) \leq 1 - \frac{C}{\binom{n}{t}^2 c^{\sqrt{t}} t!}\leq 1 - \frac{C}{n^t c^{\sqrt{t}} t!}
\]
If $t = o(n)$, this is asymptotically non-trivial.

Next, suppose that $A$ is such that $\deg(\ones_A) < \deg(\ones_B)$.
For instance, suppose that $A$ is a linear combination of $k$-juntas and $B$ is a linear combination of $m$-juntas with $k < m$.
Our lower bound states that for this case, we have
\[
\Lambda^+(f, A) \geq \left(1 - \frac{c'(m-k-1)}{(n-k)!}\right)
\]
If $m \gg k$ and $k = o(1)$, this shows that the deviation is very large.
We are effectively taking advantage of the fact that the support of an element that is a $k$-juntas is generically larger than the support of a $m$-junta, so the average over a $k$-junta's support set will be smaller than the maximum, as the maxima is contained in $A \cap B$.

\subsection{Implications for MEV}
The upper bound of Claim~\ref{claim1} can be interpreted as stating that random selection of a payoff, by first drawing $\pi \sim \mathsf{Unif}(A)$ and then realizing $f(\pi)$, can achieve some constant fraction of the maximal MEV revenue.
On the other hand, the lower bound states that the random sampling procedure cannot provide a non-trivial constant approximation of the maximal payoff if $t < s$.
The key here is that the permutations in $A$ agree sufficiently (measured by the $t$-intersection property relative to the complexity of the function $f$, measured by $s$).
As mentioned earlier, we also claim that this can be thought of as a Nyquist-Shannon sampling limit for MEV.

To see this, note that we view the Fourier decomposition of $f : S_n \rightarrow \reals$ as having `low frequency modes' corresponding to payoffs returns to sets of permutations that don't have a large overlap and `high frequency modes' as payoffs returning to sets of permutations that overlap substantially.
The bounds of the claims state that one needs sets of permutations that have sufficient overlap to realize optimal payoffs for high-degree payoff functions.

As we illustrated via our examples, CFMMs and AMMs realize low-degree payoffs, so they don't need complex sequencing rules to achieve high welfare.
This matches the results of~\cite{xavier2023credible}, where it is demonstrated that simple sequencing rules can reduce sandwich attack profitability.
On the other hand, liquidations in DeFi protocols can have very high-degree payoffs as they represent arbitrary indicator functions on sets $A \subset S_n$.
This means that random orderings impact liquidations much more and that sequencing rules to preserve different notions of fairness for liquidations much be sufficiently complex.

Our results suggest strongly that consensus enforced ordering rules need to be constructed on an application-level basis.
Each application can be viewed as represented by a payoff $f : S_n \rightarrow \reals$ and the boolean degree of $f$ controls the complexity of sequence rules needed.

\subsection{Implications for Fair Ordering}
As mentioned in~\S\ref{sec:t-intersecting}, most fair ordering protocols generate $t$-intersecting sets of permutations.
Our results suggest that fair ordering protocols that only guarantee $t = o(n)$ are likely to cause increased unfairness for payoff functions $f$ with $\deg(f) = \Omega(n)$.
This intuitively makes sense as there can be up to $O\left(\frac{n}{t}\right)$ Condorcet cycles in such cases and most of these cycles may avoid particular low payoff permutations.
On the other hand, the upper bounds of Claim~\ref{claim1} suggest that if fair ordering protocols (or general sequencing rules, such as those of~\cite{xavier2023credible}) are able to consistently generate high intersection numbers (\eg~$t = \Theta(n)$), then they are fair for `most' payoff functions.

\section{Conclusion}\label{conclusion}
In this paper, we construct a generic model for analyzing MEV under reordering.
We first formulated the problem as a discrete harmonic analysis problem on payoffs $f : S_n \rightarrow \reals$.
We then used Fourier analytic tools to relate the overlap properties of sets of permutations to different measures of fairness of $f$.
We demonstrate that if the sets of permutations do not overlap sufficiently for the payoff function $f$, then one can inure worse unfairness.
On the other hand, we showed that with sufficient overlap, the unfairness can be reduced asymptotically.

This suggests extending the works of~\cite{kulkarni2022towards, xavier2023credible}, which shows that simple sequencing rules for constant function market makers (CFMMs) can provide better fairness properties.
Since CFMMs have relatively simple and smooth payoff functions, they tend to have low-degree Fourier spectra.
On the other hand, liquidations within DeFi can have arbitrarily high-degree Fourier expansions.
This implies that sequencing rules need to be tailored to the applications that they are used with.

From a pure mathematical perspective, it is likely that our bounds can be improved.
We note that our theoretical bounds can be improved using techniques such as those of~\cite{ellis2011intersecting, keller2023t} for analyzing $t$-intersecting permutations on particular Cayley graphs.
Moreover, there is a close connection between the eigenvalue bounds used in the proof of Claim~\ref{claim:claim2} and the Aldous ordering conjecture~\cite{alon2013ordering}.
The Aldous ordering conjecture effectively states that the eigenvalues of particular representations can be bounded and/or totally ordered by `simple' representations.
We suspect that resolutions to higher order Aldous ordering conjectures will provide much stronger versions of Claim~\ref{claim:claim2}.

Finally, we note that one can extend our results to censorship MEV, related to the addition or censoring of particular transactions.
As the Young tableaux have natural decompositions into lower dimensional representations of $n$, one can effectively project the non-zero coefficients down to representations that are product representations. 
For instance, if we have a maximum of $n$ transactions in a block (\eg~maximum gas limit) then we can consider all of the decompositions $\lambda \vdash n$ and then consider the fairness functional as a function $f : S_{\lambda_1} \times \cdots \times S_{\lambda_k} \rightarrow \reals$.
However, this exponentially increases the complexity of the analysis since one has to look at the payoff function over all such decompositions.

\section{Acknowledgments}
The author would like to thank Kshitij Kulkarni, Guillermo Angeris, Peteris Erins, and Matheus V. X. Ferreira for helpful comments and suggestions.

\bibliographystyle{alpha}
\bibliography{citations.bib}

\appendix
\section{Proof of Claim 1}\label{app:indicator}
Recall that~\cite[Thm. 1]{keller2023t} proves that if $|A| = \Omega((n-t)!)$ and $c_0 t \leq n$, then there exists $i_1, \ldots, i_t$ such that $\pi(i_k) = \pi'(i_k)$ for all $\pi, \pi' \in A$.
Let $\pi \in A$ and suppose that the cycle decomposition of $A$ is $A = C_1 \cdots C_{\ell}$.
If each $i_k$ is in a different cycle $C_{\ell}$, then $\ell \geq t$.
This is because disjoint cycles don't have overlapping elements and so the Young tableaux has at least $t$ boxes as one goes downwards.
This implies that $\lambda_1 \leq n-t$ which implies that $\deg(\ones_A) \geq t$.

\section{Proof of Claim~\ref{claim1}}\label{proof-claim-1}
The majority of this proof relies on bounding $k(n, s)$, as defined in~\S\ref{sec:results}.
By definition, we have $k(n, s) = \frac{\Vert \hat{f} \Vert_1^{(S)}}{\Vert \hat{f} \Vert_{\infty}^{(S)}}$.
Recall that $\hat{f}$ takes in an irreducible representation $\rho$ of $G$ and outputs a linear operator over the same domain as $\rho$.
In particular, we have
\[
\hat{f}(\rho) = \sum_{\pi \in S_n} f(\pi) \rho(\pi)
\]
Furthermore, recall that the irreducible representations of $S_n$ are classified by the Young tableaux $\lambda = (\lambda_1, \ldots, \lambda_k)$, $\lambda_1 \geq \lambda_2 \geq \cdots \geq \lambda_k$ that are partitions of $n$ (\ie~$\sum_i \lambda_i = n$).
We will denote an irreducible representation of $S_n$ by $\rho^{\lambda} : S_n \rightarrow V^{\lambda}$ where $d_{\lambda} = \dim V^{\lambda}$.
For a matrix $A$, we will denote its singular values as $\mu_1(A) \geq \ldots \geq \mu_d(A)$.
This means that we can write the Schatten $p$-norm of a matrix $A$ as
\[
\Vert A \Vert_p^{(S)} = \Ttr[(A^*A)^{p/2}]^{1/p} = \left(\sum_{i=1}^d |\mu_i(A)|^p \right)^{1/p}
\]
Using this definition of the Schatten $p$-norm and the fact that $\hat{f}$ is only supported on partitions with $\lambda_1 \geq n - s$, we have
\begin{align*}
    \Vert \hat{f}(\rho^{\lambda}) \Vert_1^{(S)} &= \sum_{\substack{\lambda \vdash n \\ \lambda_1 \geq n-s}} d_{\lambda} \sum_{i=1}^{d_{\lambda}} |\mu_i(\hat{f}(\rho^{\lambda}))| \\
    \Vert \hat{f}(\rho^{\lambda}) \Vert_{\infty}^{(S)} &= \max_{i \in [d_{\lambda}]}  |\mu_i(\hat{f}(\rho^{\lambda}))|
\end{align*}
Since $t \geq s$, we have $n-t \leq n-s$ so that this expression makes sense for $\pi \in A$.
Now note the following upper bound:
\begin{align*}
\frac{\Vert \hat{f}(\rho^{\lambda}) \Vert_1^{(S)}}{\Vert \hat{f}(\rho^{\lambda}) \Vert_{\infty}^{(S)} } &= \frac{\sum_{\substack{\lambda \vdash n \\ \lambda_1 \geq n-s}} d_{\lambda} \sum_{i=1}^{d_{\lambda}} |\mu_i(\hat{f}(\rho^{\lambda}))|}{\max_{i \in [d_{\lambda}]}  |\mu_i(\hat{f}(\rho^{\lambda}))|} = \sum_{\substack{\lambda \vdash n \\ \lambda_1 \geq n-s}} d_{\lambda} \sum_{i=1}^{d_{\lambda}} \frac{|\mu_i(\hat{f}(\rho^{\lambda}))|}{\max_{i \in [d_{\lambda}]}  |\mu_i(\hat{f}(\rho^{\lambda}))|} \\
&\leq \sum_{\substack{\lambda \vdash n \\ \lambda_1 \geq n-s}} d_{\lambda}^2
\end{align*}
One can show an upper bound on $d_{\lambda}$ of the following form~\cite[Pg. 136]{diaconis1988group}:
\begin{equation}\label{eq:diaconis}
d_{\lambda} \leq \binom{n}{\lambda_1} \sqrt{(n-\lambda_1)!}
\end{equation}
Moreover, the number of partitions $\lambda \vdash n$ that have $\lambda_1 = n-i$ is precisely the number of partitions of $i$.
Recall that number of partitions of $k$ has asymptotics of the form $O(k^{-1}c^{\sqrt{k}})$ for $c > 1$.
Therefore we have
\begin{align*}
k(n, s) \leq \sum_{\substack{\lambda \vdash n \\ \lambda_1 \geq n-s}} d_{\lambda}^2 &\leq \sum_{\substack{\lambda \vdash n \\ \lambda_1 \geq n-s}} \binom{n}{\lambda_1}^2 (n-\lambda_1)! \leq D \sum_{i=1}^s \frac{c^{\sqrt{i}}}{i} \binom{n}{i}^2 i! \leq D c^{\sqrt{s}} \binom{n}{s}^2 s! 
\end{align*}
for some $D > 0$.
This proves the result.

\section{Proof of Claim 2}\label{app:proof-of-claim-2}
We first provide two key results that we will use to get our final result.
The first is a symmetric group analogue of the so-called `second' Erd\"os-Ko-Rado theorem.
\begin{theorem}[\cite{keller2023t}]
There exists a universal constant $c_0 > 0$ such that for all $t \in \naturals$ and $n \geq c_0 t$ and a t-intersecting set $A \subset S_n$, we have $|A| \leq (n-t)!$. Moreover, if $|A| = \Omega((n-t)!)$, then there exist $(i_1, j_1), \ldots, (i_t, j_t)$ such that for all $\pi \in A$, $\pi(i_k) = j_k$.
\end{theorem}
We note that this result was first conjectured in the 1960s and proved by~\cite{ellis2011intersecting} for $t = O(\log \log n)$.
The second result required comes from~\cite[\S7]{filmus2020hypercontractivity}.
Recall that a symmetric set $F \subset S_n$ is a subset such that if $\pi \in F$ then $\pi^{-1} \in F$.
Given a symmetric set, we can construct the Cayley graph $G = (S_n, E)$ on $S_n$ where $\pi, \pi'$ are connected by an edge (\ie~$(\pi, \pi') \in E)$ if there exists $\pi'' \in F$ such that $\pi = \pi'' \circ \pi'$.
Let $T_F$ be the adjacency matrix of the this graph given a symmetric set $F$.
A simple result (\cite[Claim 7.5]{filmus2020hypercontractivity}) is that $T_F$ maps the space of functions $V^{\lambda}$ to itself.
Note, that $T_F$ as an adjacency matrix can be lifted to an operator on functions $f : S_n \rightarrow \reals$ such that
\[
T_F(f(\pi)) = \Expect_{(\pi, \sigma) \in E}[f(\sigma)]
\]
The eigenvalues of $T^*_F T_F$, $\mu_{\lambda}$, for each $\lambda \vdash n$ satisfy~\cite[Claim 7.6]{filmus2020hypercontractivity}
\begin{equation}\label{eq:eigenvalues}
\mu_{\lambda} \leq \frac{n!}{|F|\dim(\lambda)}
\end{equation}

Now we are ready to prove the claim.
Let $\hat{A} = A \cup \{a^{-1} : a \in A\}$ be the symmetrized version of $A$.
Note that $T_{\hat{A}} = \frac{1}{2}(T_A + T_{A}^t)$.
Let $g = (f^{\leq s})^{\geq t} = f - f^{\leq t}$.
Firstly, note the following elementary inequality due to Claim~\ref{claim:indicator-degree}
\[
\Vert g \ones_A\Vert_1 \leq \Vert g \ones_A \Vert_{\infty} \Vert \ones_{A}^{\leq t} - \ones_{A}^{\geq s}\Vert_1 
\]
The comes from factoring out the maximal term when writing out the 1-norm and the fact that $g$ is only supported on $\lambda$ with $\lambda_1 \in [n-s, n-t-1]$.
We claim that
\[
\Vert \ones_A^{\leq s} - \ones^{\geq t}_A \Vert_1 \leq \sum_{\substack{\lambda \vdash n \\ \lambda_1 \in [n-s, n-t-1]}} \mu_{\lambda}
\]
where $\mu_{\lambda}$ are the eigenvalues of $T_{\hat{A}}$.
Recall that $T_F V^{=\lambda} \subseteq V^{=\lambda}$ so if we write out
\[
(\ones_A^{\leq s} - \ones^{\geq t}_A)(\pi) = \sum_{\substack{\lambda \vdash n \\ \lambda_1 \in [n-s, n-t-1]}} \ones_A^{=\lambda}
\]
which then implies that 
\[
\Vert T^*_{\hat{A}} T_{\hat{A}}(\ones_A^{\leq s} - \ones^{\geq t}_A)\Vert_1 = \Vert\sum_{\substack{\lambda \vdash n \\ \lambda_1 \in [n-s, n-t-1]}}   T^*_{\hat{A}}T_{\hat{A}}\ones_A^{=\lambda}\Vert_1 \leq \sum_{\substack{\lambda \vdash n \\ \lambda_1 \in [n-s, n-t-1]}}  \Vert  T^*_{\hat{A}} T_{\hat{A}}\ones_A^{=\lambda}\Vert_1 = \sum_{\substack{\lambda \vdash n \\ \lambda_2 \in [n-s, n-t-1]}} 
 \mu_{\lambda}
\]
Note that $\Vert  T^*_{\hat{A}} T_{\hat{A}}(\ones_A^{\leq s} - \ones^{\geq t}_A)\Vert_1  \geq \Vert \ones_A^{\leq s} - \ones^{\geq t}_A\Vert_1$ since $T_{\hat{A}}$ is non-contractive by~\eqref{eq:eigenvalues}.
Now note that by assumption, $|A| \geq C(n-t)!$ so that $\mu_{\lambda} \leq \frac{(n)_t}{\dim(\lambda)}$.
Note that via the hook length formula, $\dim(\lambda) \geq c^n$ for some $c > 1$.
This gives us the final bound:
\begin{align*}
    \Vert g \ones_A\Vert_1 &\leq \Vert g \ones_A \Vert_{\infty} \Vert \ones_{A}^{\leq t} - \ones_{A}^{\geq s}\Vert_1 \leq \Vert g \ones_A \Vert_{\infty} \sum_{\substack{\lambda \vdash n \\ \lambda_2 \in [n-s, n-t-1]}} \mu_{\lambda} \\
    &\leq \Vert g \ones_A \Vert_{\infty} \sum_{\substack{\lambda \vdash n \\ \lambda_1 \in [n-s, n-t-1]}} \frac{(n)_t}{c^n} 
    \\ 
    &= \Vert g \ones_A \Vert_{\infty} (n)_t \left(\sum_{\substack{\lambda \vdash n \\ \lambda_1 \in [n-s, n-t-1]}} \frac{1}{c^n}\right) \\
    &=  \Vert g \ones_A \Vert_{\infty} (n)_t \left(\sum_{i = t+1}^s \frac{c^{\sqrt{i}}}{s} \frac{1}{c^n} \right)
\end{align*}
We note that since $ic^n \geq c^{\sqrt{i}}$ for $i \in [n]$ the sum is bounded by $D(s-t-1)$ for a constant $D \geq 0$.
Thus we have:
\[
\Expect[g\ones_A] = \frac{1}{n!}\Vert g\ones_A \Vert_1 \leq \frac{\Vert g \ones_A \Vert_{\infty} (n)_t}{n!} D(s-t-1)
\]
as claimed

\end{document}